\documentclass[aps,10pt,twocolumn]{revtex4-1}
\usepackage{hyperref}
\usepackage{enumerate}
\usepackage{amsmath}
\usepackage{graphicx}
\usepackage{graphics}
\usepackage{dcolumn}%
\usepackage{bm}%
\usepackage{color}
\usepackage{subfigure}
\usepackage{float}
\usepackage{docs}%
 \usepackage[]{lineno} 
\begin{document}

\title{ Equilibrium configuration of self-gravitating charged dust clouds: Particle approach } 

\author{Manish K. Shukla}
\email[Electronic mail : shuklamanish786@gmail.com]\
\author{K. Avinash}
\affiliation{Department of Physics and Astrophysics, University of Delhi, Delhi, 110007, India}

\begin{abstract}  
A three dimensional  Molecular Dynamics (MD) simulation is carried out to explore the equilibrium configurations of charged dust particles. These equilibrium configuration are of astrophysical significance for the conditions of molecular clouds and the interstellar medium. The interaction among the dust grains is modeled by Yukawa repulsion and gravitational attraction. The spherically symmetric equilibria are constructed which are characterized characterized by three parameters: (i) the number of particles in the cloud, (ii)  $\Gamma_g$ (defined in the text) where $\Gamma_g^{-1}$ is the short range cutoff of the interparticle potential, and (iii) the temperature of the grains. The effects of these parameters on dust cloud are investigated using radial density profile. The problem of equilibrium is also formulated in the mean field limit where total dust pressure which is the sum of kinetic pressure and electrostatic pressure, balances the self-gravity. The mean field solutions agree well with the results of  MD simulations. Astrophysical significance of the results is briefly discussed.

\end{abstract}

\keywords{ Dusty plasma, Molecular dynamics, Equations of state, Self-gravity, Jeans instability, Interstellar cloud }

\pacs{52.25.Kn, 52.27.Lw, 52.65.Yy, 98.38.Dq, 98.38.-j }

\maketitle

\section{Introduction\label{Sec. 0}}
Dust is one of the main constituents of the universe and plays a central role in the formation of stars and planets\cite{harpaz1993stellar,desch2004astr, hartmann1983moons}. A variety of astrophysical systems such as interstellar clouds, solar system, planetary rings, cometary tails, Earth's environment etc. contain micron to sub-micron sized dust particles in different amounts \cite{frank2000waves}. However, in the presence of ambient plasma and radiative environments, these particles are acted upon by the electron-ion currents, photons and energetic particles. As a result, the dust grains acquire a non-negligible electric charge (which is usually negative) and start interacting with each other, as well as with ions and electrons via long-range electric fields giving rise to collective behavior. This leads to the formation of special medium called the ``dusty plasma''. 

The macroscopic particles in  dusty plasmas  interact not only via electric field but also via gravitation field which is also a long ranged force.  Which of the forces dominates, is determined by the ratio of average gravitational potential energy to average electrostatic (ES) potential energy. If $m_d$, $Q_d$ be the mass, charge of a dust grain and $n_d$ be the number density then the gravitational potential $\psi$ is given by Poisson's Equation, $\nabla^2 \psi=4 \pi G m_d n_d $ which scales as $\psi \approx 4\pi  Gm_dn_d L^2$, therefore, average gravitational energy of a dust grain may be given as $m_d\psi \approx 4\pi Gm_d^2n_d L^2 ,$ where  $G$ is the universal gravitational constant and $L$ is the typical length scale over which dust is distributed. Similarly, the ES potential $\phi$ for dust is given  as  $\nabla^2 \phi=-Q_d n_d/\epsilon_0 $, thus, $\phi$ scales as $\phi \approx -Q_dn_dL^2/\epsilon_0$ and average ES potential energy scales as $Q_d \phi \approx Q_d^2n_dL^2/\epsilon_0$ and the ratio of two energies comes out to $G m_d^2/(Q_d^2/4\pi\epsilon_0)$.  For typical plasmas of size of order of microns and charge of the order of $10^3 \; e$, the ratio comes out to be $ \approx \mathcal{O}(10^{-17}) $.  However, in dusty plasmas, the electric field is screened substantially by the presence of the background electrons and ions. The estimate of ES potential, in such case, should be made using the quasi-neutrality condition $Q_d n_d = e(n_i-n_e)$. Taking the electron and ion response as Boltzmannian  i.e. $n_\alpha=n_0\exp(q_\alpha\phi/k_BT_\alpha)$ where $\alpha=$(electron, ion), $T_\alpha$ is the temperature of spices $\alpha$ and $n_0$ is the mean plasma density in the region where $\phi=0$, the ES energy  scales as  $Q_d \phi \approx Q_d^2 n_d\lambda_D^2/\epsilon_0$, where $\lambda_D= \left({e^2n_0(T_e+T_i)}/{\epsilon_0k_B T_e T_i}\right)^{-1/2}$ is the Debye screening length. Therefore, the ratio of gravitational energy to electrostatic energy in the presence of background plasma becomes $G m_d^2 L^2/(Q_d^2\lambda_D^2/4\pi\epsilon_0)$. Since, the typical length scale of dust cloud, $L\gg \lambda_D$ therefore, the ratio $G m_d^2 L^2/(Q_d^2 \lambda_D^2/4\pi\epsilon_0)$ could be $\approx \mathcal{O}(1)$. Therefore, for the  micron sized particles the forces due to screened ES fields and gravitational force are comparable on sufficiently long length scales $L$ which may be found in the astrophysical scenarios. 

The Jeans instability\cite{jeans1929} is the most fundamental instability in the self-gravitating systems which plays a crucial role in the formation of structures in the universe. Physically, whenever there is an imbalance between the self-gravitating force and the internal pressure of a gas, the Jeans instability appears. The threshold for the Jeans instability in self-gravitating neutral fluid is set by the hydrostatic pressure of the fluid\cite{chandrasekhar1961hydrodynamic}. In case of dusty plasmas, the neutral mass density is replaced by the dust mass density and the neutral gas pressure is replaced by the sum of electron and ion pressure. 
 
Therefore, the threshold for Jeans instability is set by the dust acoustic waves\cite{rao1990dust} instead of the usual sound waves\cite{Shukla_Stenflo,Pandey_avinash1994, avi_shukla_1994, PhysRevE.60.7412, pandey_dwivedi_1996, RAO2000}. Jeans instability in  quantum dusty plasmas is studied by a number of authors \cite{SHUKLA2006378, MASOOD20086757, doi:10.1063/1.3070664, doi:10.1063/1.3168612} and the effect of physical processes like magnetic field, radiative cooling, polarization, electron-ion recombination and dust charge fluctuations etc. on Jeans frequency is reported  in Refs. [\onlinecite{prajapati2010effect,doi:10.1063/1.4961914,1742-6596-836-1-012029}]. One major problem encountered frequently in dealing with the self-gravitating matter is the formation of legitimate equilibrium. On the astrophysical scales, the dust kinetic pressure is too weak to balance the self-gravity which suggests that the dust clouds are gravitationally unstable.

Recently, Avinash and Shukla have shown a new way of forming stable dust structures by balancing the force due to self-gravity of the dust\cite{avi2006_masslimit,avi_Eliasson_shukla2006}. In the case of dusty plasmas, the equilibrium configuration of dust cloud (embedded in a background plasmas) can be constructed by balancing the the forces due to screened ES field and gravitational forces which as pointed out earlier, are comparable on sufficiently longer length scales. The electrostatic repulsion among dust grains is equivalent to an ``electrostatic pressure''\cite{avi2006_electrostatic_pressure}. If dust gains are distributed inhomogeneously within the plasma background, the electrostatic pressure, like the kinetic pressure, expels the dust from the regions of high dust density to the regions of low dust density. Avinash and Shukla have also shown the existence of a mass limit for the total mass supported by ES pressure against gravity \cite{avi2006_masslimit}. 
At low dust density the ES pressure $P_{ES}$ scales quadratically with number density $n_d$ ($ P_{ES} \propto n_d^{\gamma}~, \gamma=2$). 
At very  high dust densities ES pressure becomes independent of number density due to charge reduction caused by mutual screening of the grains\cite{Avinash2003,Barkan} and as a consequence $\gamma \rightarrow 0$. This change of $\gamma$ from two to zero implies that with increasing dust density, $P_{ES}$ is not able to cope with the gravitation resulting an upper mass limit ($M_{AS}$). The physics of this mass limit is very similar to Chandrasekhar's mass limit for white dwarfs where adiabatic index for hydrostatic pressure changes from $5/3$ to $4/3$ due to relativistic effects.  If the mass of dust cloud exceeds this upper mass ($M> M_{AS}$) the ES pressure is not strong enough to balance the gravity and the system will undergo symmetric collapse under self-gravity \cite {avi2006_masslimit,avi_Eliasson_shukla2006,avinash_2010}. Using the energy principle, the stability of these charge clouds is formulated and it is established that collapse responsible for $M>M_{AS}$ is due to stability of radial eigen mode \cite{avi2007_equilibrium,Avinash_energy_principle}. 

We, in this communication, revisit the equilibrium problem using molecular dynamics (MD) simulations. The particles are subject to Yukawa repulsion as well as gravitation attraction force. At the equilibrium of these forces, a spherically symmetric cloud is formed. 
We also probe the dust density distribution of these clouds using radial density profile. The dependence of the equilibrium structure on the number of particles $N_d$, dimensionless parameters $\Gamma_g$  and temperature  of the grains is examined using radial density profile. 
Unlike the previous models \cite{avi2006_masslimit,avi_Eliasson_shukla2006,avi2007_equilibrium, BORAH201549} which are based on fluid description, our approach is based on particle description. One can go from particle description to fluid description via proper ensemble averaging, a process commonly known as ``coarse graining". We have, therefore, also formulated the problem of equilibrium in the mean field limit in terms of force balance coupled 
with equation of state. The mean field solution are are compared with the MD results and the the two solutions are found to agree well with each other.

The equilibrium structures are of considerable interest because of their astrophysically relevant length and mass scales. It is well known that the dust particles of micron and sub-micron size are an important component of interstellar clouds\cite{turner1989nature}. 
This dust, which is typically silicate or polycyclic aromatic hydrocarbon, is the dominant cause of opacity and reddening of the spectral energy distribution of radiation from distant stars\cite{evans1993dusty,krugel2002physics}. The HI and HII regions are the part of interstellar molecular clouds where hydrogen is found to exist in abundance along with other gases like He, CO and the dust grains. In the HI region, which  are relatively cooler with temperature $\sim 100$ K, hydrogen is found in the neutral state. The HII regions which are spread over the size ranging from 0.01 pc to 10 pc (1 pc $=3.08\times 10^{16} $ m), are the part of molecular clouds where star formation has recently taken place. The HII regions are relatively hotter with temperature ranging from $5000\--10000$ K and hydrogen is found in the ionized state in this region.  
The ionization takes place mainly due to the UV radiation of the nearby stats. In this conducting medium the macroscopic dust grains pick the negative charge from the medium$\--$ typically due to the attachment of negative electrons on the dust surface. The charge on the dust surface depends on the plasma conditions and its magnitude is about $10^2\--10^4~e$ , where $e$ is the electronic charge. The typical mass of a micron sized dust is about $10^{10}\--10^{13}$ times the mass of proton. The high resolution data from Herschel space observatory\cite{anderson2012dust} suggest that the dust is very cold in the HII region and the typical range of dust temperature varies from $\sim 12 \-- 40$ K. 

In this scenario, the interstellar dust first undergo gravitational instability. This instability saturates when the electrostatic pressure becomes equal to the gravitation force density which results in formation of a tenuous dust clouds which are stable. In the later section of this paper, we will show that for the physical parameter of HII region, the equilibrium structure formed in this process have approximately same order of size as the size of clumps of clouds observed\cite{braun2005tiny,smith2012small} in the interstellar medium. The density profile explored in this paper not only gives the important information about the matter and charge distributions inside the clouds but may also provide an important insight about optical thickness of the small scale structures observed in the interstellar medium. 

The paper is organized in the following manner. In Sec. \ref{Sec.I} the details of the model used and the assumptions made in the model are described briefly.  In Sec. \ref{Sec. II}, we describe the details of the interaction potential, normalization used and the methodology used for MD simulations.
The simulation results are given in Sec. \ref{Sec. III}. The mean field solutions are derived in Sec. \ref{Sec. IV} and validation of simulation results with mean field solution is provided in Sec. \ref{Sec. V}. The summary of the work is given in  Sec. \ref{Sec. VI} where we have also have discussed the astrophysical significance of the present.

\section{Model \label{Sec.I}}
Our model consist of a finite sized dust particles embedded in a much larger background of warm hydrogen plasma consisting of electrons and protons. Such a situation could occur in the HII region, where dust clouds have been observed\cite{fischer2012hubble} in the background of hydrogen plasma. 

In the interstellar medium dust grains exhibit a varying range of size, mass and charge. However, we make a simplifying assumption that all the dust grains have same size and therefore, equal mass. The electron and ion are much lighter than dust grains i.e. $m_d \gg m_i,~ m_e$ and plasma temperature $ T_e\approx T_i=T \gg T_d$.  The charge on the dust particle, unlike the intrinsic charge on electron and ions, depends on the local plasma environment. The important mechanism which are responsible for the charging of the dust grains in the interstellar medium are the thermal flux of electrons and ions and the flux of the photo-electrons \cite{spitzer1968diffuse,wickramasinghe2012theory}. The estimate of the dust charge in such case, can be made using orbital motion limited theory \cite{shukla2015introduction}. We, in our model, assume that the charge on all the dust particle due to various physical processes is equal with constant magnitude $Q_d$. It should be noted here that the dust size plays an important role in determining dust mass and dust charge. If $r_d$ is the radius the dust particle then the dust mass $m_d\propto r_d^3$ whereas dust charge  $Q_d\propto r_d$. However, in our simulation once we assign the mass and charge of the dust particles then size has no direct role to play in the equilibrium formation. Therefore, in our simulation, we can safely consider the dust particle as point particle with mass $m_d$ and charge $Q_d$. 

In our model, we assume that only the massive dust particles are affected by the gravity and their response in MD simulation is taken into account by using Newton's law of gravitation. The background plasma is considered to be unaffected by the gravity. This situation is likely to occur in the HII region where $T_e=T_i=T\approx 5000$ K and plasma density, $n_0 \approx 10^7~ m^{-3}$. For these parameters, the gravitational potential energy of plasma is less than their thermal energy i.e., $GM_cm_i/R_c \ll k_BT$ (where $m_i$ is the ion mass, $M_c$ and $R_c$ are respectively the mass and radius of the dust cloud). Therefore, the gravitational collapse of the plasma background may be neglected and we may consider a thermalized, static plasma background governed by the Boltzmann relation for electrons and ions, i.e., $n_e = n_0 \exp (e\phi/k_BT), ~ n_i = n_0 \exp(͑-e\phi/k_BT)$. Here $\phi$ is the local plasma potential and $\phi=0$ at infinity where dust density is zero i.e.,  $n_d=0$ and $n_e=n_i=n_0$. This is a reasonable assumption for the interstellar clouds\cite{spitzer1968diffuse,wickramasinghe2012theory}. 
The electrostatic interaction among dust grains embedded in the  background of statistically averaged neutralizing plasma background is given by the screened Coulomb (or Yukawa) potential.

In the presence of self-gravity, the dust-dust interaction becomes,
\begin{equation}
 \Phi(r) =   \frac{Q_d^2}{4\pi\epsilon_0}\frac{e^{-r/\lambda_D}}{r} -  \frac{Gm_d^2}{r}, 
 \label{eq. 1}
\end{equation}
which is the sum of Yukawa and gravitational potential energy. Here, $r$ is the distance between two dust particles and $\lambda_D$ is the screening length defined earlier. It is well known that thermodynamics of gravitating systems has problems due to absence of short range and long ranged cutoffs in the inter-particle potential. However, in self gravitating dusty plasmas the short range cutoff is naturally provided by the electrostatic repulsion and the long range cutoff is defined by the typical length scale of dust dispersion. Hence, thermal equilibria of self gravitating dusty plasma exist at all temperatures. 

\section{Simulation Details \label{Sec. II}}


\begin{figure*}[]
\begin{center}
 \mbox{\subfigure[][]{\includegraphics[scale=0.65]{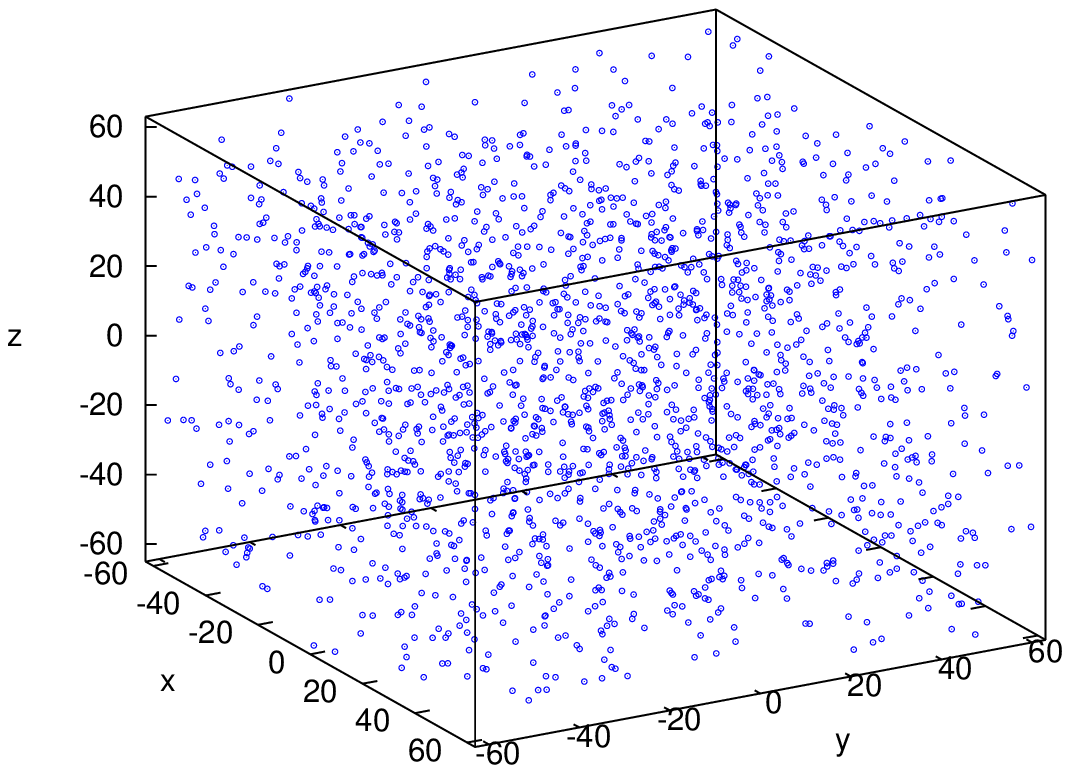}\label{fig:1a}}
   \subfigure[][]{\includegraphics[scale=0.55]{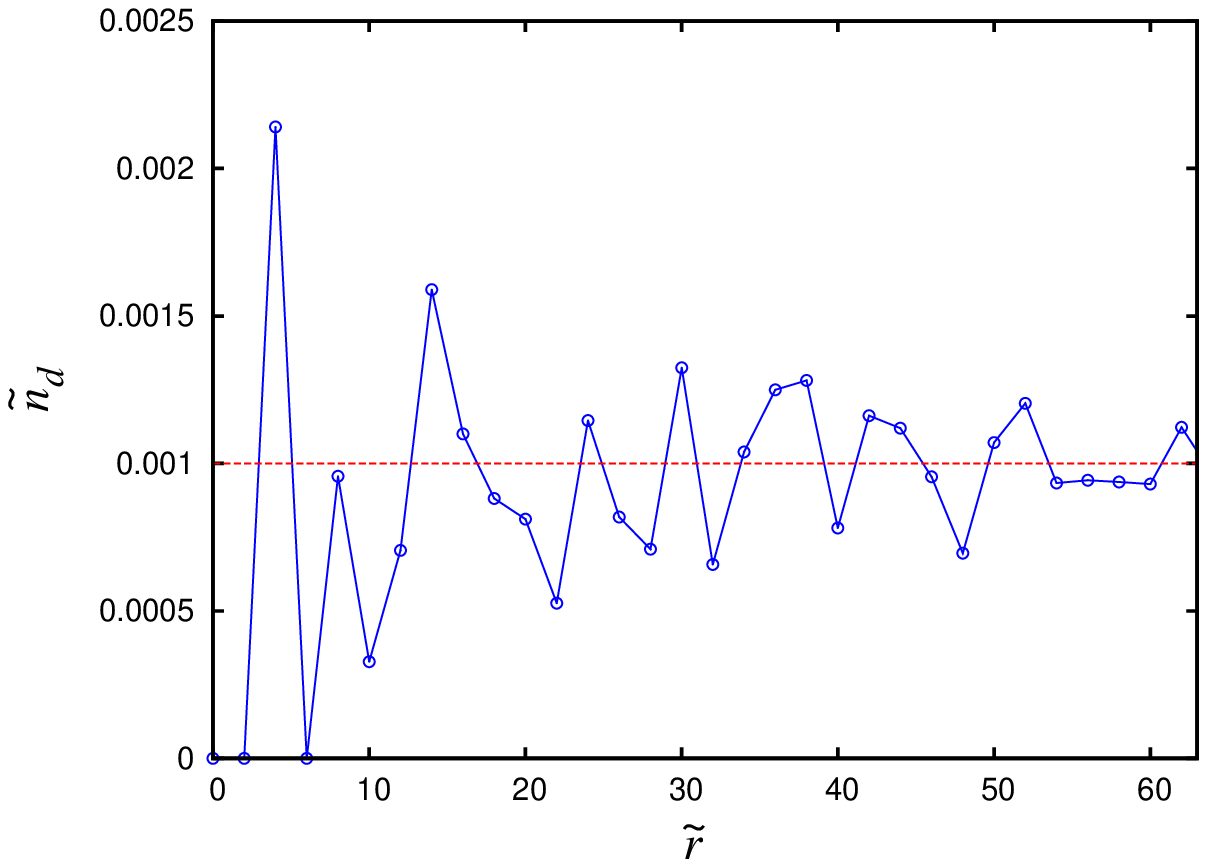}\label{fig:1b} } }
 \mbox{\subfigure[][]{\includegraphics[scale=0.65]{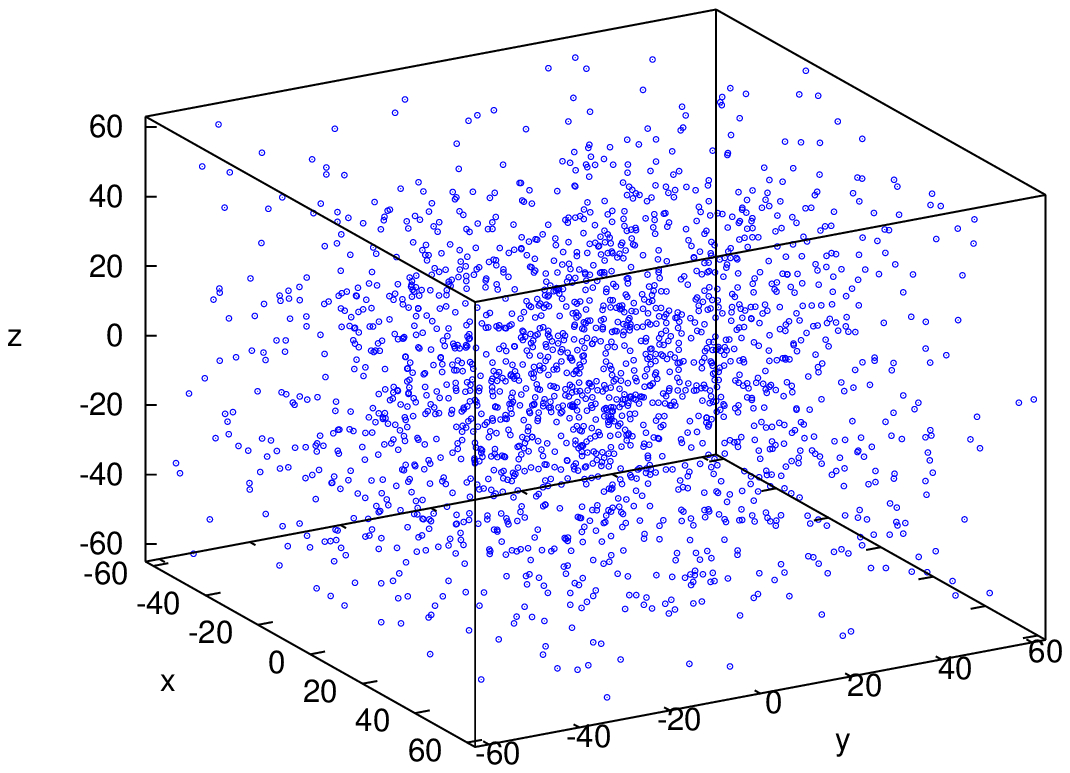}\label{fig:1c}}
   \subfigure[][]{\includegraphics[scale=0.55]{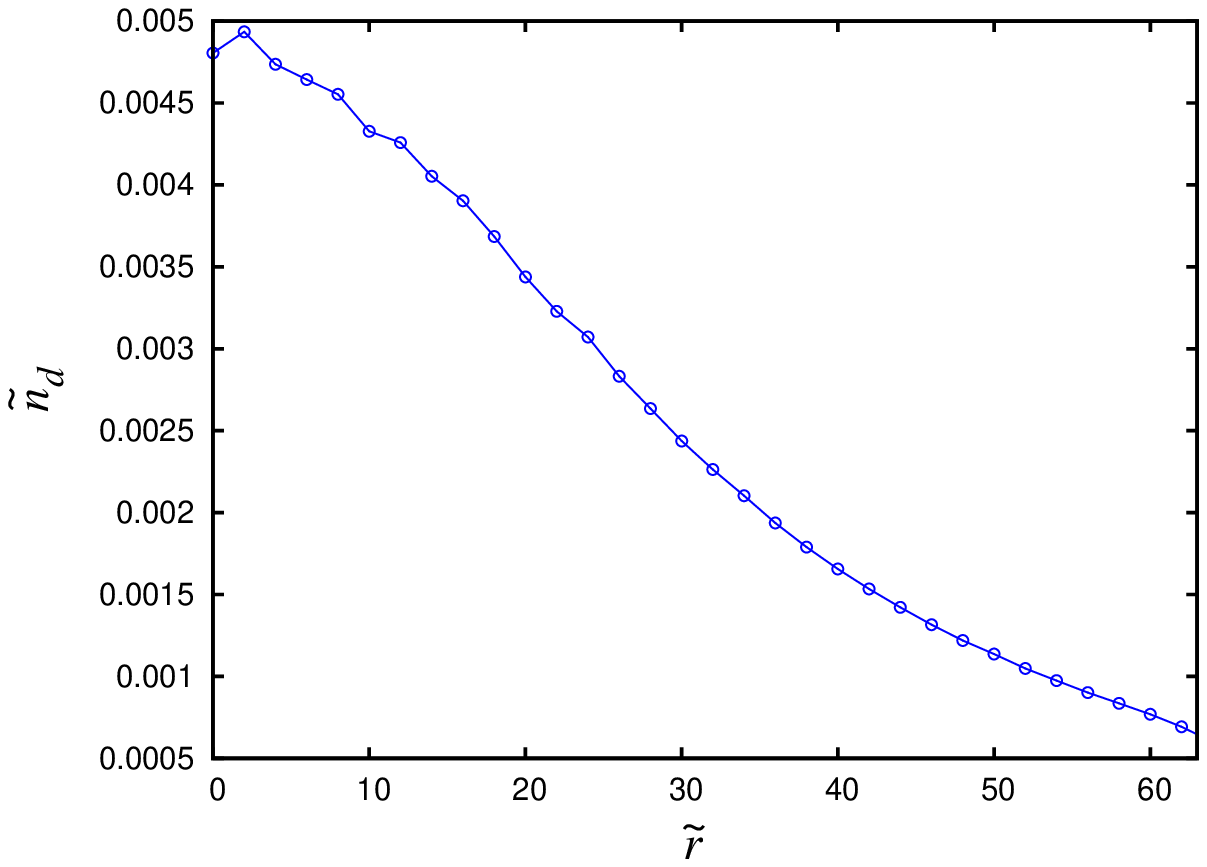}\label {fig:1d}}}  
  \mbox{\subfigure[][]{\includegraphics[scale=0.65]{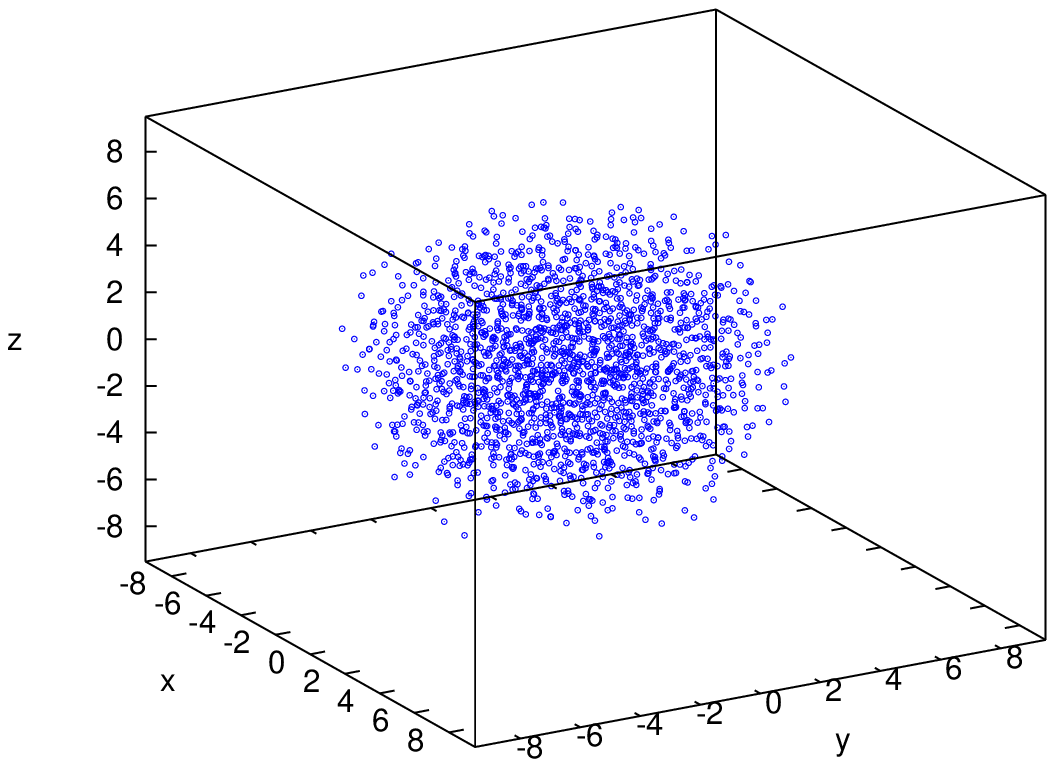}\label{fig:1e}}
   \subfigure[][]{\includegraphics[scale=0.55]{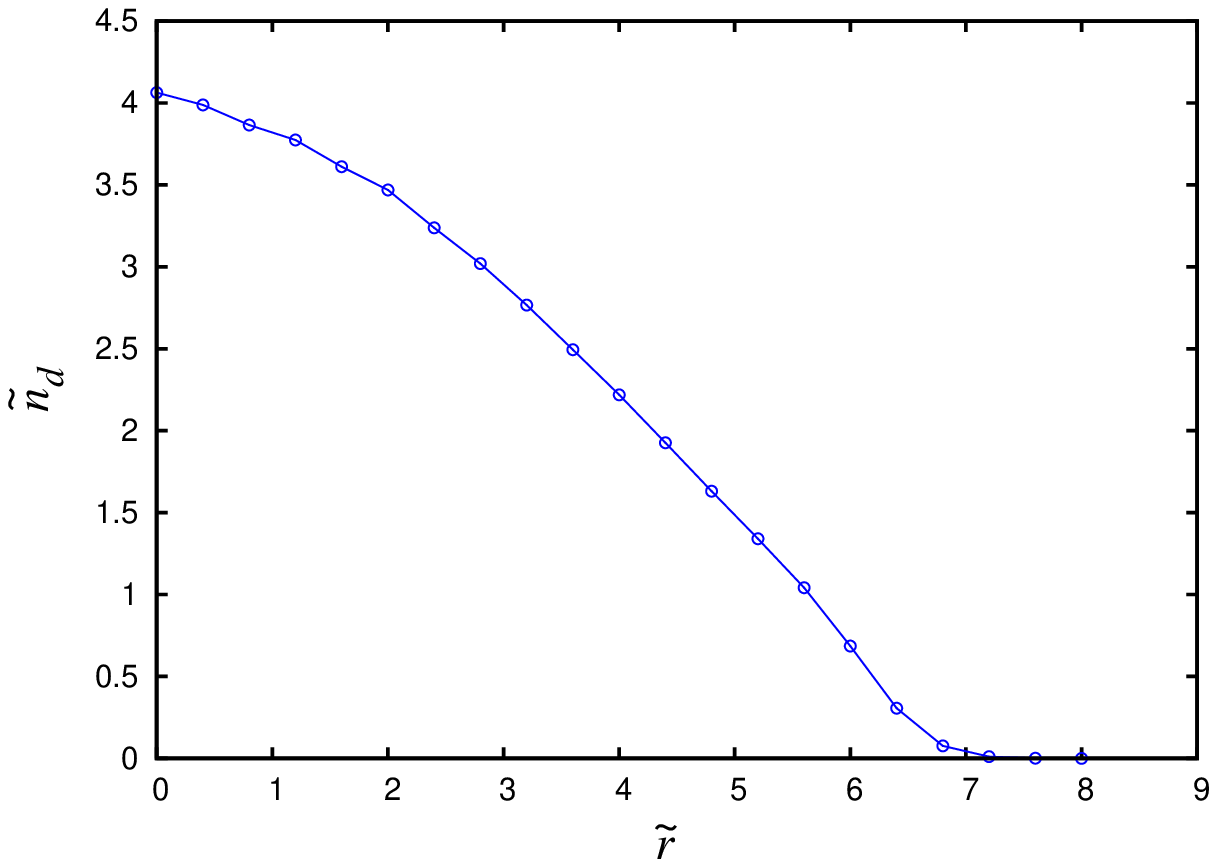}\label {fig:1f}}}  
  \caption{  Different states of self-gravitating Yukawa particles is shown for  $N_d=2000$. At the initial state particles are uniformly distributed in a cubical box with number density $\tilde n_d =0.001$ as shown in Fig.\ref{fig:1a} and corresponding radial density distribution is shown in Fig.\ref{fig:1b}. Fig.\ref{fig:1c} refers to the equilibrium structure of the system where dust temperature $\tilde T_d=1.5$ and the corresponding density profile is shown in Fig.\ref{fig:1d}.  At lower temperature $\tilde T_d=0.5$ the equilibrium structure acquires the shape of sphere with a very dense core as shown in Fig.\ref{fig:1e}. The averaged radial density of such dust cloud is shown in Fig.\ref{fig:1f}. }
  \label{fig:1}
 \end{center}
\end{figure*} 


\subsection{Equations and Units \label{subsec.IIA}}

The Hamiltonian of the system is given by;
\begin{equation}
 H=\sum_{i=1}^{N_d} \frac{p_i^2}{2m_d} +  \frac{1}{2}  \sum_{i=1}^{N_d}\sum_{j,j\neq i}^{N_d} \Phi(r_{ij})
 \label{eq. 2}
\end{equation}
where $r_{ij}=  |\vec{r_i}-\vec{r_j}|$ and $p_i=|\vec p_i|$. We have assumed that all grains have equal charge $Q_d$ and mass $m_d$. The equation of motion are given by 
\begin{equation}
 \frac{d\vec r_i}{dt}= \frac{\vec {p}_i}{m_d}, \\
 \frac{d\vec p_i}{dt}= -\vec{\nabla}_{r_i} \left(  \sum_{i=1}^{N_d}\sum_{j>i}^{N_d}  \Phi(r_{ij}) \right),
 \label{eq. 3}
\end{equation}
All the lengths are normalized by the  background Debye length $\lambda_D$ and time is measured in units of $\omega_0^{-1}$ where $\omega_0 = \left( Q_d^2/4\pi \epsilon_0 m_d \lambda_D^3\right)^{1/2}$. Hence, 
$ r \rightarrow \tilde{r}\lambda_D $  and $t \rightarrow \tilde{t} \omega_0^{-1}$, where $\tilde{r}$ and $\tilde{t}$ are dimensionless length and time respectively. Therefore, dimensionless equation of motion is
\begin{equation}
 \frac{d \vec{\tilde {r}}_i}{d\tilde t}= \vec{\tilde {v}}_i \;, \:
 \frac{d \vec{\tilde {v}}_i}{d\tilde t}=\vec{\tilde f}_i 
 \label{eq. 4}
\end{equation}
where
\begin{equation}
 \vec{\tilde{f}}_i=\sum_{j,j\neq i}^{N_d} 
 \left((1+ \tilde r_{ij}) \frac{e^{-\tilde r_{ij}}}{ \tilde r_{ij}^3}-\frac{ (4\pi\epsilon_0 Gm_d^2/Q_d^2)}{\tilde r_{ij}^3} \right)  \vec{ \tilde r}_{ij},
 \label{eq. 5}
\end{equation}
Temperature ($k_B T$) and energy is measured in units of $Q_d^2/4\pi\epsilon_0\lambda_D$ i.e. $\tilde{E}=E/(\frac{Q_d^2}{4\pi\epsilon_0 \lambda_D})$ and $\tilde{T}_d= k_BT_d/(\frac{Q_d^2}{4\pi\epsilon_0\lambda_D})$.
The dimensionless kinetic and potential energy per particles can be written as 
\begin{eqnarray}
 \tilde{E}_K &=&\frac{1}{2N_d} \sum_{i=1}^{N_d} \tilde{v}_{i}^2 ~,\\
 \tilde{E}_{Pot}&=&\frac{1}{2N_d}\sum_i^{N_d} \sum_{j, j\ne i}^{N_d} \frac{(e^{-\tilde{r}_{ij}} -  \Gamma_g) } {\tilde{r}_{ij}} ~,
\end{eqnarray}
where $\Gamma_g$ is a dimensionless parameter defined as,  
\begin{equation}
 \Gamma_g=\frac{Gm_d^2}{(Q_d^2/4\pi\epsilon_0 )}.
 \label{eq. 6}
\end{equation}
In our simulations $\Gamma_g^{-1}$ defines the short range cutoff in inter-particle potential. In addition to this, $\Gamma_g$ also contains important information of various parameters of interest. As $\Gamma_g$ is directly proportional to $m_d^2/Q_d^2$, the effect of the dust mass, dust charge and dust size ($\Gamma_g\propto r_d^4$) can directly be seen from $\Gamma_g$.

Temperature is given by the mean kinetic energy per particle, which, in dimensionless units in 3D, turns out to be
$$\tilde{T}_d = \frac{1}{3 N_d}\sum_{i=1}^{N_d} \tilde{v}_{i}^2=\frac{2}{3} E_K.$$
Normalized number density is given as, $$\tilde{n}_d = n_d \lambda_{D}^3 .$$ 

Dusty plasma equilibrium is defined in terms of two dimensionless  parameters: $\kappa=a/\lambda_D $ and $\Gamma={Q_d^2}/{4\pi\epsilon_0 a T_d} $ where $a$ is the mean inter-particle distance given by $ a=\left( 3/4\pi n_d\right)^{1/3} $.  The coupling parameter $\Gamma^{*}= \Gamma \exp(-\kappa)$ which is the ratio of the mean inter-particle potential energy to the mean kinetic energy, is used as a measure of coupling strength in dusty plasmas. For $\Gamma^* \ll 1$, Yukawa system behaves like an ideal gas, $\Gamma^* \sim 1$ corresponds to an interacting fluid whereas $\Gamma^* \gg 1$  refers to a condensed solid state. The dimensional parameters $\kappa$ and $\Gamma$ can be related to normalized dust density and temperature as;
\begin{equation}
 \kappa= \left(\frac{3}{4\pi\tilde{n}_d}\right)^{1/3}, \hspace{0.5cm} \Gamma = \frac{1}{\kappa \;\tilde{T}_d}.
 \label{Gamma-kappa}
\end{equation}
Instead of choosing ($\Gamma$, $\kappa$) space to work in, we prefer to work in ($\tilde n_d$, $\tilde T_d$) space, however, one can switch from one space to other space  using the relations given in Eq.(\ref{Gamma-kappa}).

\subsection{Methodology \label{subsec.IIB}}

All the simulation are performed using a large scale OpenMP parallel Three Dimensional Molecular Dynamics (3DMD) code developed by authors. 
The further details regarding the units, equations and  methodology are given in the following Secs. \ref{subsec.IIA}-\ref{subsec.IIB}.  

Simulation begins with an initial condition where $N_d$ number of particles are distributed randomly over a volume such that the initial number density is $\tilde n_d = 10^{-3}$. The equation of motion, given in Eq.(\ref{eq. 4}), is integrated taking time step of size $0.01\; \omega_0^{-1}$ . This step size is appropriate for the conservation of total energy (i.e. sum of kinetic and potential energy).
The system is kept in contact with a heat bath using Berendsen thermostat for the fist $2\times 10^5$ steps which bring the system at desired temperature after which the system is isolated and it remains isolated for another $2\times 10^5$ steps. For those runs where  the temperature  $\tilde T_d \ge 1.0$, the step size is halved and the total number of steps is doubled to achieve numerical accuracy. All the observations are taken  during the micro-canonical run when the system is isolated. As the system evolves, depending upon $\Gamma_g$, $N_d$ and $\tilde T_d $, different equilibrium structures are obtained as shown in Fig.(\ref{fig:1}). To get the information about the interior of the equilibrium structure we plot spherically symmetric radial density distribution $\tilde n_d(r)$. While plotting the $\tilde n_d$ as a function of $\tilde r$, the radial distance is measured from the center of mass of the cloud. The final radial distribution is the ensemble average of hundreds of individual copies of $\tilde n_d(\tilde r)$ taken at different time intervals. 

In the next section we discuss the results of our simulations.


\begin{figure}[h]
 \includegraphics[scale=0.70]{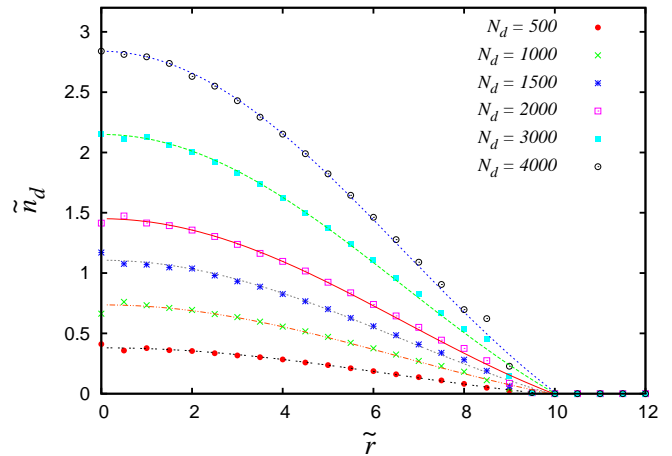}
   \caption{  The effect of number of particles on the radial  density is shown for the fixed value of $\Gamma_g=0.08$ and $\tilde T_d=0.10$. The symbols represent the data points from MD simulation whereas the curves represent the best fit of the corresponding data using trial function of the form given in right hand side of  Eq.(\ref{eq. 16}). It can be seen from the plot that the radius of dust cloud is almost independent of $N_d$ and any increases in the number of particle only increases the density of the core. }
   \label{fig.2}
\end{figure}


\begin{figure}[h]
 \includegraphics[width=0.99\linewidth]{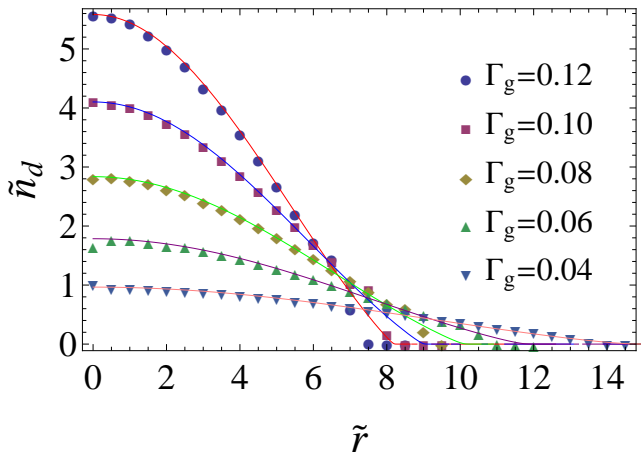}
\caption{ Effect of $\Gamma_g$ on the radial density profile: Radial density profile $\tilde n_d$ is plotted against the radial distance $\tilde r$ for different values of $\Gamma_g$ for the fixed $N_d=4000$ and $T_d=0.10$. The symbols represent the density profile obtained  from the MD simulation whereas the solid lines correspond to the mean field solution of $\tilde n_d(\tilde r)$ obtained from Eq.(\ref{eq. 13}).  It is clear from the results that the radius of the cloud $\tilde R$ decreases whereas the core density $\tilde n_{d0}$ increases with $\Gamma_g$. }
\label{fig:3}
\end{figure}

\begin{figure}[ht]
\begin{center}
 \includegraphics[width=0.95\linewidth]{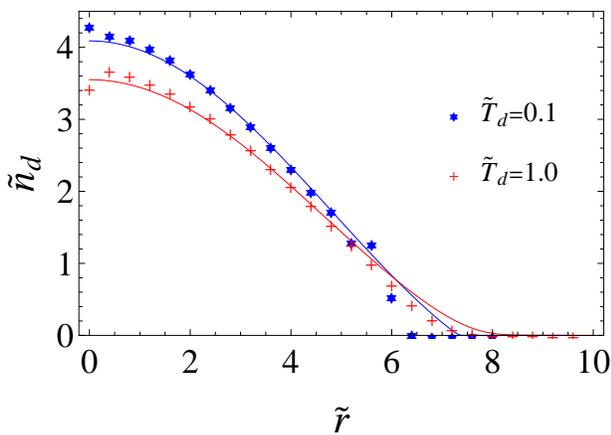} 
  \caption{ Effect of dust temperature on radial number density is shown for $\Gamma_g=0.15$ and $ N_d =2000 $  for $\tilde T_d=0.1$ and  $\tilde T_d=1.0$. Symbols represent data points from the simulation whereas curves refer to the mean field solutions. 
}
  \label{fig.4} 
 \end{center}
\end{figure} 

\section{Simulation Results \label{Sec. III}}
The size of the cloud critically depends on three parameters,  $N_d$,  $\Gamma_g$ and $\tilde T_d$. 
Following these observations, the simulation results may broadly be divided into three parts accordingly; one, comprising the effect of $N_d$ on $\tilde n_d(\tilde r)$ keeping $\Gamma_g$ and $\tilde T_d$ fixed, second, the effect of  $\Gamma_g$ on $\tilde n_d$ for fixed $N_d$ and $\tilde T_d$ and third, the effect of $\tilde T_d$ on $\tilde n_d(\tilde r)$ keeping $\Gamma_g$ and $N_d$ fixed.

\subsection{Effect of $N_d$ on density profile \label{Sec.III A}}
To examine the effect of $N_d$ of $\tilde n_d(\tilde r)$, we fix $\Gamma_g =0.08$ and $\tilde T_d = 0.10$. The equilibrium radial density profile of dust cloud for different $N_d$ is shown by symbols in Fig.(\ref{fig.2}). It can be seen that as we  increase the number of particles $N_d$ the central density (i.e. the  density at $\tilde r=0$) increases whereas the radius of the cloud $\tilde R$ (value of $\tilde r$ at which $\tilde n_d =0$ ) is almost fixed. In other words, the dust cloud becomes denser with increase in the number of particle while size of the cloud is almost independent of  the number of particles.

\subsection{Effect of $\Gamma_g$ on density profile\label{Sec.III B}}
The effect of $\Gamma_g$ on radial number density $\tilde n_d(\tilde r)$ for fixed number of particles $N_d=4000$ and constant temperature $T_d=0.10$ is shown in Fig.(\ref{fig:3}) where symbols represent the simulation data points and curves correspond to the mean field solution discussed in next section. It is  clear from the  Fig.(\ref{fig:3}), that the radial density, $\tilde n_d(\tilde r)$, radius of the cloud, $\tilde R$ and central density, $\tilde n_{d0}$, depend on $\Gamma_g$. 

\subsection{Effect of dust temperature on density profile \label{Sec.IV C} }
Effect of dust temperature on the radial density profile for $N_d=2000$ and $\Gamma_g=0.15$ is shown  in Fig.(\ref{fig.4}).
The equilibrium number density is plotted against radial distance for $\tilde T_d=0.1$ and  $\tilde T_d=1.0$. It is seen that the central density of cloud  decreases while radius  increases with increase in temperature. 

The observed simulation results can be explained in terms of mean field theory\cite{avi2006_masslimit,avi2007_equilibrium} described in the next section.

\section{Mean Field Solutions \label{Sec. IV}}
In the mean field (continuum) limit, the electric force acting on the dust fluid behave like an effective ES pressure force. It can be seen as follows: In the zero correlation mean field limit, $Q_d \rightarrow 0$, $N_d \rightarrow \infty $ and $Q_dN_d=\text{finite}$, therefore, the double summation can be changed to smooth integration\cite{avinash2010thermodynamics} as $\sum_{i}^{N_d} \sum_{\hspace{.1cm}j=1,\neq i}^{N_d} \rightarrow n_d N_d \int_{V_d}  \; dV_d $. 
In this limit ES energy becomes;
$$
U_{ES}=  \frac{Q_d^2}{4\pi\epsilon_0} \frac{1}{2} \hspace{-.1cm} \sum_{i}^{N_d}\hspace{-.2cm} \sum_{\hspace{.1cm}j=1,\neq i}^{N_d} \hspace{-.2cm}\frac{\exp(-\kappa_D|r_i -r_j|)}{|r_i-r_j|} \rightarrow \frac{n_dN_d Q_d^2 \lambda_D^2}{2\epsilon_0}. 
$$
Corresponding ES pressure can be obtained using the relation $ P_{ES}=-\left({\partial U_{ES}}/{\partial V_d} \right)_{T_d}$ which comes out to be 
$P_{ES}= {Q_d^2\lambda_D^2  n_d^2}/{2 \epsilon_0}$ suggesting that $P_{ES}\propto n_d^2$.  The total dust pressure is the sum of kinetic pressure ($n_dT_d$) and ES pressure\cite{mks2017}. Similarly the  gravitational force density, in the mean field limit, is given by $-m_dn_d \nabla \psi $ where $\psi$ is gravitational potential  given by Poisson equation  i.e.,
\begin{equation}
 \nabla^2 \psi= 4 \pi G \rho_d.
 \label{eq. 9}
\end{equation} 
Therefore, equation of motion for such self-gravitating  dust fluid is given by
\begin{equation}
 \rho_d \frac{du_d}{dt}= -\nabla P_d -\rho_d \nabla \psi , 
 \label{eq. 7}
\end{equation}
where $\rho_d(=m_dn_d)$ is dust mass density, $P_d$ is total dust pressure (sum of kinetic pressure and ES pressure). In static equilibrium $u_d=0$,  therefore 
\begin{equation}
 \nabla P_d =-\rho_d \nabla \psi .
 \label{eq. 8}
\end{equation} 
Taking the divergence of Eq.(\ref{eq. 8}) (after dividing by $\rho_d$) and using Eq.(\ref{eq. 9}), we get
\begin{eqnarray}
 \nabla \cdot \left(\frac{ \nabla P_d}{\rho_d}\right) &=&   \nabla^2 \psi  \nonumber \\
  &=& 4\pi G \rho_d,
  \label{eq. *}
\end{eqnarray}
The spherically symmetric force balance equation in the spherical polar coordinated becomes; 
\begin{equation}
 \frac{1}{r^2} \frac{d}{dr} \left(\frac{r^2}{m_d n_d} \frac{dP_d}{dr}\right)= -4 \pi G m_d n_d,
 \label{eq. 10}
\end{equation}
where we still have to provide an equation of state for $P_d$ to close Eq. (\ref{eq. 10}).

As our purpose is to compare the simulation results with the mean field solution, let us normalize Eq.(\ref{eq. 10}) as we did in Sec.(\ref{Sec. II}) where $ r \rightarrow \tilde{r}\lambda_D $, $P_d \rightarrow \tilde{P}_{d} (\frac{Q_d^2}{4\pi\epsilon_0 \lambda_D^4})$ and  $n_d \rightarrow \tilde n_d/ \lambda_{D}^3 $. Therefore Eq. (\ref{eq. 10}) becomes,
\begin{eqnarray}
\frac{1}{\tilde r^2} \frac{d}{d\tilde r} \left(\frac{\tilde r^2}{ \tilde n_d} \frac{d\tilde P_d}{d\tilde r}\right) &=& -\frac{4 \pi G m_d^2}{Q_d^2/4 \pi \epsilon_0} \tilde n_d  \nonumber \\
&=& -4 \pi \Gamma_g \tilde n_d.
 \label{eq. 11}
\end{eqnarray} 
Recently, Shukla \emph{et al} \cite{mks2017} have obtained the equation of state for Yukawa fluid  by using rigorous MD simulations.
Their expression for total dust pressure in normalized units is  given by
\begin{equation}
 \tilde{P}_d =  \tilde{n}_d \tilde{T}_d + \beta \: \tilde{n}_d^{2},
 \label{eq. 12}
\end{equation}
where $\tilde T_d$ is dust temperature and $\beta$ is a number of the order of $\pi$. In the expression of $\tilde P_d$, the first term which scales linearly with number density, is usual kinetic pressure term whereas second term which is proportional to the square of number density, corresponds to ES pressure.
Substituting Eq.(\ref{eq. 12}) in Eq.(\ref{eq. 11}), we get the following differential equation
\begin{equation}
 \tilde n_d''+\frac{\tilde T_d}{2\beta}\frac{\tilde n_d''}{\tilde n_d} +\frac{2}{r} \tilde n_d'+ \frac{\tilde T_d}{2\beta} \frac{2}{r} \frac{\tilde n_d'}{\tilde n_d} - \frac{\tilde T_d}{2\beta}  \frac{(\tilde n_d')^2}{\tilde n_d^2}= -\frac{2\pi \Gamma_g}{\beta} \tilde n_d
\label{eq. 13}
 \end{equation}
where $\tilde n_d' = d\tilde n_d/d\tilde r$ and $\tilde n_d'' = d^2\tilde n_d/d\tilde r^2$. 
As Eq.(\ref{eq. 13}) is a second order nonlinear differential equation, it requires two boundary conditions (BCs) for a unique solution which are given by;
\begin{equation}
 \tilde n_d(0) = \tilde n_{d0}, \hspace{0.45cm} \tilde n_d'(0)= 0. 
 \label{eq. 14}
\end{equation}
Eq.(\ref{eq. 13}) along with Eq.(\ref{eq. 14}) may be solved numerically to get $\tilde n_d(\tilde r)$ for a given temperature $\tilde T_d$, central density $\tilde n_{d0}$ and $\Gamma_g$. 
The radius of the cloud ($\tilde R$) is calculated using the relation $\tilde n_d(\tilde R)=0$. 
whereas the mass of the cloud is obtained  using the relation 
\begin{equation}
 M_d=m_dN_d=4\pi m_d\int_0^R \tilde r^2 \tilde n_d(\tilde r) d\tilde r.
\end{equation}

Before comparing the MD results with the mean field solutions, we consider the special case of zero dust temperature in which Eq.(\ref{eq. 13}) admits an exact analytic solution.

\subsection*{Special Case: Solution for $\mathbf{\tilde T_d=0}$}
There is an interesting case in the limit $\tilde T_d=0$ when the  Eq.(\ref{eq. 13}) can be solved analytically\cite{avi2006_masslimit}.  In this special case, the gravitation force is balanced completely by ES pressure force.  In this limit  Eq.(\ref{eq. 13}) reduces to a linear differential equation given by; 
\begin{equation}
 \tilde n_d^{''} +\frac{2}{r} \tilde n_d^{'} = -\frac{2\pi \Gamma_g}{\beta} \tilde n_d. 
\label{eq. 15}
 \end{equation}
The solution of above differential equation along with  BCs defined in Eq.(\ref{eq. 14}), is given by
\begin{equation}
 \tilde n_d(\tilde r)= \tilde n_{d0} \frac{\sin\left(\sqrt{\frac{2\pi \Gamma_g}{\beta} } \;\tilde r \right)}{\left(\sqrt{\frac{2\pi \Gamma_g}{\beta}} \;\tilde r \right)}.
 \label{eq. 16}
\end{equation}
As mentioned earlier the radius of the cloud is obtained by setting $\tilde n_d(\tilde R)=0$, which gives $\sqrt{\frac{2\pi \Gamma_g}{\beta} } \;R=\pi $. Therefore,  radius of cloud is; 
\begin{equation}
 \tilde R= \sqrt {\frac{\pi \beta}{2 \Gamma_g}}.
 \label{eq. 17}
\end{equation}
This should be noted that this radius is of the same order as obtained by equating average gravitational field to mean ES field discussed in Sec. \ref{Sec. 0} which gives the dust Jeans length ($L$ also provides the typical length scale of dust dispersion) as $L\sim \lambda_D \sqrt{1/\Gamma_g}$. 

Relation between central density $\tilde n_d$ and $N_d$ can be obtained by integrating number density as follows
\begin{equation}
 N_d= \int_0^{\tilde R} \tilde n_d(\tilde r) 4\pi \tilde r^2 d\tilde r.
 \label{eq. 18}
\end{equation}

Substituting the value of $\tilde n_d$ and $\tilde R$ from Eq.(\ref{eq. 16}) and (\ref{eq. 17}) respectively in above equation gives,

\begin{equation}
 \tilde n_{d0}= \frac{1}{\sqrt{2\pi}} \left(\frac{\Gamma_g}{\beta} \right)^{3/2} N_d
 \label{eq. 19}
\end{equation}
It should also be noted here that in the limit of $\tilde T_d \rightarrow 0$, the radius of the cloud $\tilde R$  is independent of the number of particles $N_d$ and is function of $\Gamma_g$ only [Eq.(\ref{eq. 17})] whereas $\tilde n_{d0}$ is directly proportion to $N_d$ and   $\Gamma_g^{3/2}$ [Eq.(\ref{eq. 19})]. 

\section{Validation of simulation with mean field solutions \label{Sec. V}}

The particle approach can be tested against the fluid approach by comparing MD results with the mean field solutions. Let's compare them one by one.

\begin{figure}[ht]
\begin{center}
 \subfigure[][]{\includegraphics[scale=0.70]{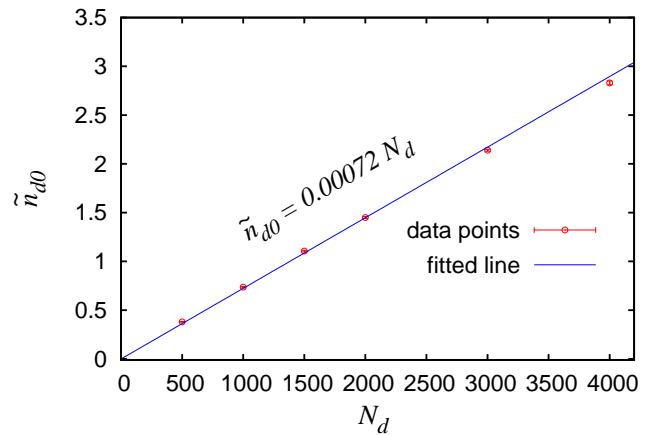}\label{Fig. a}}
 \subfigure[][]{\includegraphics[scale=0.70]{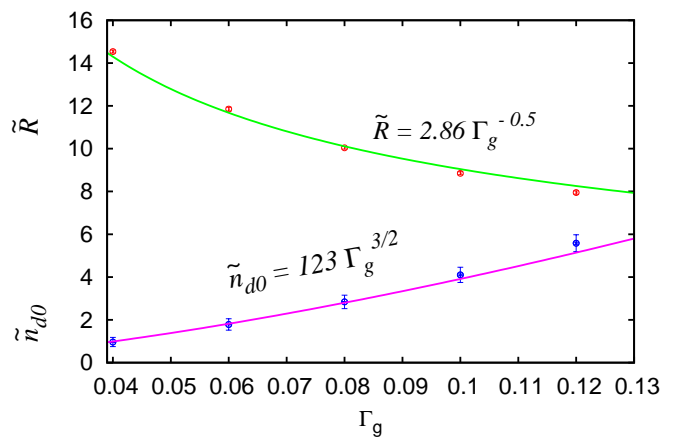}\label{Fig. b}}  
  \caption{ The density at the center of cloud, $\tilde n_{d0}$, is plotted against number of particles, $N_d$ in Fig.\ref{Fig. a}.  The $\tilde n_{d0}$ vs. $N_d$ plot is linear in nature and is consistent with Eq.(\ref{eq. 19}). The dependence of $\tilde n_{d0}$ and size on parameters $\Gamma_g$ is shown in the Fig. \ref{Fig. b}.  }
  \label{fig:5} 
 \end{center}
\end{figure}

In Sec. [\ref{Sec.III A}] we have observed  that the increase in $N_d$ does not affects the radius of cloud substantially but increases the density of cloud. A similar relation is seen in mean field solution in the limit $\tilde T_d \rightarrow 0$ through Eq. (\ref{eq. 17}) and Eq. (\ref{eq. 19}). Therefore, we  fit our MD data for radial density profile in a trial function of the form given in Eq. (\ref{eq. 16}),
$$ \tilde n(r)=\tilde n_{d0} ~\frac{\sin (kr)} {(kr)}, $$ where parameters $\tilde n_{d0}$ and $k$ are obtained by least square fitting and are tabulated in Table. \ref{table. 1}. The fitted curves are plotted  along with simulation data in Fig. (\ref{fig.2}). It is clear from Table. \ref{table. 1} that value of $k$, which in mean field limit represents to $\sqrt{2\pi\Gamma_g/\beta}$, is constant under the numerical errors and so does the the radius of the cloud.  
Also, the constancy of $k$ or $\sqrt{2\pi\Gamma_g/\beta}\approx 0.315\pm 0.006$  fixes $\beta \approx 1.604\; \pi$.  
The  Eq.(\ref{eq. 19}) states  that the central density  $\tilde n_{d0}$ should be a linear function of $N_d$ and the predicted slope is $\frac{1}{\sqrt{2\pi}} \left(\frac{\Gamma_g}{\beta} \right)^{3/2}$ which comes out to be $  7.98 \times 10^{-4}$ for $\Gamma_g =0.08$ and $\beta = 1.604\; \pi $. To verify Eq.(\ref{eq. 19}), we fit  central density vs. number of particles using  data from  Table \ref{table. 1}. The slope of $\tilde n_{d0}$ vs. $N_d$ line comes out to be  $(7.23\pm0.07) \times 10^{-4} $ against the predicted value of $ 7.98 \times 10^{-4}$ thereby validating the results with a deviation of about $10\%$ (Fig (\ref{Fig. a})).
\begin{table}[h]
 \caption{Equilibrium parameters for different $N_d$ }
\begin{tabular}{c c c c}
\hline 
$~~\mathbf{N_d}~~$ & $~~~~\tilde n_{d0}~~~~~~$ & $~~~~~~$k$~~$ &$~~\tilde R=\pi/k~~$\\
\hline
500  & $0.380\pm0.003$ & $0.321\pm 0.003$ & $9.78 \pm0.09$\\
1000 & $0.736\pm0.008$ & $0.316\pm 0.003$ & $9.94 \pm0.09$\\
1500 & $1.106\pm0.009$ & $0.317\pm 0.003$ & $9.91 \pm 0.09$\\
2000 & $1.45\pm 0.01$  & $0.315\pm 0.002$ & $9.97 \pm0.06$\\
3000 & $2.14\pm 0.01$  & $0.313\pm 0.002$ & $10.03\pm 0.06$\\
4000 & $2.83\pm 0.02$  & $0.312\pm 0.002$ & $10.03\pm 0.06$ \\ 
\hline
\end{tabular}
\label{table. 1}
\end{table}


The fitted parameters for different values of $\Gamma_g$ keeping $N_d$ and $T_d$ constant, are tabulated in Table \ref{table. 2}. In the mean limit, the relation between $\tilde R$ and $\Gamma_g$, using Eq.(\ref{eq. 17}), reduces to $\tilde R=2.813\; \Gamma_g^{-1/2}$ while relation between $\tilde n_{d0}$ and $\Gamma_g$, as predicted by Eq.(\ref{eq. 19}), is given by $\tilde n_{d0}= 141.07\; \Gamma_g^{3/2}$ (taking $N_d=4000$ and $\beta=1.60\; \pi$).
A direct fit between $\Gamma_g$ and $\tilde R$ using Table \ref{table. 2} gives the relation $\tilde R=2.86\pm0.02 \;\Gamma_g^{-1/2}$ while relation between $\tilde n_{d0}$ and $\Gamma_g$ comes out to be $\tilde n_{d0}= 123\pm1 \; \Gamma_g^{3/2}$ (Fig.(\ref{Fig. b})). 

It should be noted that the parameters $\tilde n_{d0}$, $k$ shown in Table \ref{table. 1} and Table \ref{table. 2} are obtained by fitting the radial density data in the function $\tilde n(r)=\tilde n_{d0} \sin (kr)/(kr)$, which represents the mean field solution in the limit $\tilde T_d\rightarrow0$. It is also noted that the simulation results at $T_d=0.1$ is a good approximation for analytic solutions at $T_d=0$. In Fig. (\ref{fig:3}) we plot the mean field solutions (solid curves) for different $\Gamma_g$ obtained by solving Eq.(\ref{eq. 13}) with  $\tilde T_d=0.10$. Simulation results agree well with mean field solutions for different $\Gamma_g$.    

\begin{table}[h]
 \caption{Equilibrium parameters for different $\Gamma_g$}
\begin{tabular}{c c c c}
\hline 
$~~\mathbf{\Gamma_g}~~$ & $~~~~\tilde n_{d0}~~~~~~$ & $~~~~~~k~~$ & $~~\tilde R=\pi/k~~$\\
\hline
0.04 & $0.965\pm0.005$& $ 0.216\pm0.001$ & $14.54\pm0.07$ \\
0.06 & $1.78\pm 0.01$ & $0.265\pm 0.002$ & $11.85\pm 0.09$ \\
0.08 & $2.84\pm 0.02$ & $ 0.312\pm0.002$ & $10.04\pm 0.06$ \\
0.10 & $4.10 \pm0.04$ & $0.355\pm 0.003$ & $8.85\pm 0.07$ \\
0.12 & $5.58\pm0.05$  & $0.395\pm0.004 $ & $7.95\pm0.08$ \\ 
\hline
\end{tabular}							
\label{table. 2}
\end{table}

To compare the effect of temperature in two approaches, we plot MD data (symbols) and the solution of  Eq.(\ref{eq. 13}) (curves) for different dust temperature as shown in Fig. (\ref{fig.4}). While plotting mean field solution, the central density is taken from simulation results and Eq.(\ref{eq. 13}) is solved numerically. It can be seen from the plots that the mean field field solutions are validating  simulation results. 
\begin{figure}[h]
 \includegraphics[scale=0.70]{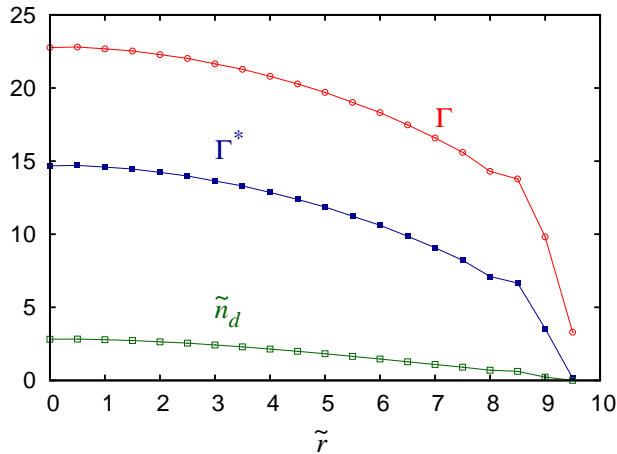}
   \caption{ Typical values of the Coulomb coupling parameter $ \Gamma $ and the coupling parameter $ \Gamma^* \left(=\Gamma \exp(-\kappa)\right)$ corresponding to radial density profile $\tilde n_d(\tilde r)$ is shown for $ N_d=4000$ and $\Gamma_g=0.080 $. }
   \label{fig.6}
\end{figure}


\section{Summary and Conclusion \label{Sec. VI}}

To summarize, we have examined the problem of equilibrium of self-gravitating dusty plasmas using particle level MD simulations. Dust grains interact with each other via  repulsive Yukawa potential and attractive gravitational potential. The equilibrium of the system is characterized by three parameters, $\Gamma_g (=4\pi\epsilon_0 G m_d^2/Q_d^2)$, number of particles $N_d$ and mean kinetic energy or temperature $\tilde T_d$ and 
depending upon these three parameters, different equilibrium structures are formed. The interior of these equilibrium structures is probed using radial density function where, center of mass of the cloud is taken as the $r=0$ point. We have also formulated the problem of equilibrium  in the mean field limit where dust pressure, which is the sum of kinetic pressure and ES pressure, balances the self gravity. The results of mean field limit  are compared with simulation results and the two approaches are found to be consistent with each other. 

Our results predict that for the cold dust particles of constant charge and mass, the size of equilibrium structure is independent of number  of particle in the cloud (or mass of the cloud). In fact, the addition of more number of particles results only  in increasing the number density (or equivalently mass density) of the equilibrium cloud. Our results also predict that the equilibrium structures formed by the dust particles of relatively bigger size will be shorter in size and denser in nature. This happens because the increase in the dust size results in the increase of  $\Gamma_g ~(\text{which is a sensitive function of dust size as,} ~\Gamma_g \propto r_d^4$) and the size of the equilibrium structure,  $R \propto \lambda_D /\sqrt{\Gamma_g}\propto 1/r_d^2. $ The increase in temperature results in increasing the radius of equilibrium cloud. The effect of dust temperature is obvious. In our model, it is the sum of kinetic pressure and electrostatic pressure which balances the self-gravity and therefore, the increase in dust temperature implies an increase in kinetic pressure which ultimately pushes the dust grains outwards from the cloud, thereby increasing the size of the cloud.

It should be noted that the equation of state for ES pressure i.e. $P \propto n_d^2 $ is derived with an assumption of weak coupling \cite{avinash_2010,mks2017}, however,  it may be very robust and  could  be valid  in the strong coupling regime. For example, the 
mean field solution matches well even near the central region of dust cloud where dust is not weakly correlated and $\Gamma^*>1$ as shown in Fig. (\ref{fig.6}). Similar quadratic scaling of dust pressure is also observed in experiments on shock formation in a flowing 2D dusty plasma where Saitou \emph{et al.} have shown\cite{saitou2012bow} that the condition of shock formation is satisfied by equation of state $P_d \propto n_d^{\gamma}$ where $\gamma \simeq2.2$. Some other examples where quadratic scaling is found to be valid in the even in the presence of correlations includes the simulations of Charan \emph{et al.} where $P_d \propto n_d^{2}$ scaling is seen in the regions where dust particles are compressed by external gravity and dust is strongly coupled\cite{charan2014properties}. The simulation of Djouder \emph{et al.} for dust monolayer confined by parabolic potential shows that the equation of state  near zero dust temperature  follows the relation $P_d \propto n_d^{2\sim2.165}$ for a wide range of densities near the dust crystal\cite{djouder2016equation}. 

We now briefly discuss the astrophysical significance of our results. The observations in the infrared region of spectrum have shown the evidences of dust \emph{overabundance} inside the HII regions as compared to the interstellar medium\cite{panagia1974infrared,tenorio1974presence}. The length scale of these over dense clumps is below 1 pc to several AUs (1 pc $=3\times 10^{16}$ m, 1 AU $=1.5\times 10^{11}$ m).The origin of these structures is not known even today. We propose that the equilibrium structures discussed in our paper could be a possible candidate for these small scale structures observed in the HII region and interstellar medium. In our model the typical length scale of these structures is given by $R  = \lambda_D/\sqrt{\Gamma_g}$. 
For the parameters of HII region: plasma density $n_0=10/$ c.c. (=$10^7 \text{m}^{-3}$), $T=5000$ K and average dust size $\approx 0 .3\;\mu$m we obtain $ R\approx 3\times 10^{11}$ m which is roughly of the size of finest structure of clumps of gas and dust detected in interstellar medium \cite{quirrenbach1989rapid, braun2005tiny,smith2012small}. The equilibrium structures discussed in this paper could also be the precursor to a proto-planetary or proto-stellar core formation. For example, if the electric fields become weak over a period of time then these aggregates will slowly contract and become denser to give rise to van der Waals correlations and the formation of a more solid body\cite{avi2006_masslimit}.
 
It should also be noted that while formulating the problem of equilibrium, dust charge is taken to be constant and independent of number density while there are enough evidences to conform that charge of dust decreases with number density \cite{Havnes,Avinash2003,Barkan}. As mentioned in Sec.\ref{Sec. 0} the reduction of dust charge at high density can limit the total mass supported by the ES pressure. Therefore our study is useful to the scenarios where dust density is low and dust charge is constant. A  number  of other  effects like  magnetic field, dust rotation etc. are also not taken into account and will be addressed in future communication.

\acknowledgments
MKS acknowledges  the financial support from University Grants Commission (UGC), India, under JRF/SRF scheme. MKS is also thankful to Prof. R. Ganesh, IPR for his kind help regarding the MD simulation and providing the access to IPR's high performance cluster machines Uday and Udbhav  on which the major part of computing of this paper has been done.

%

\end{document}